




\magnification\magstep1 
\baselineskip14pt
\vsize23.5truecm

\input miniltx
\input graphicx






\def\hatt{\widehat}
\def\dell{\partial}
\def\tilda{\widetilde}

\def\half{\hbox{$1\over2$}}

\def\arr{\rightarrow}
\def\normal{{\cal N}}

\def\RR{\mathord{I\kern-.3em R}}
\def\PP{\mathord{I\kern-.3em P}}
\def\NN{\mathord{I\kern-.3em N}}
\def\ZZ{\mathord{I\kern-.3em Z}} 
\def\Var{{\rm Var}}
\def\E{{\rm E}}
\def\d{{\rm d}}

\def\midd{{\,|\,}}
\def\subsection{\medskip}
\def\df{{\rm df}}

\font\bigbf=cmbx12

\font\csc=cmcsc10

 at 10truept 
\font\smallrm=cmr8

\def\today{\number\day \space \ifcase\month\or
January\or February\or March\or April\or May\or June\or 
July\or August\or September\or October\or November\or December\fi  
\space \number\year}


   
\def\ref#1{{\noindent\hangafter=1\hangindent=20pt
  #1\smallskip}}          

\def\quotationone{\smallrm Where there is a Will}
\def\quotationtwo{\smallrm There is a Won't}
\def\hskipdistanceleft{\hskip-3.5pt}
\def\hskipdistanceright{\hskip-2.0pt}
\footline={{
\ifodd\count0
        {\hskipdistanceleft\quotationone\phantom{\smallrm\today}
                \hfil{\rm\the\pageno}\hfil
         \phantom{\quotationone}{\smallrm\today}\hskipdistanceright}
        \else 
        {\hskipdistanceleft\quotationtwo\phantom{\today}
                \hfil{\rm\the\pageno}\hfil
         \phantom{\quotationtwo}{\smallrm\today}\hskipdistanceright}
        \fi}}

         
\def\cstok#1{\leavevmode\thinspace\hbox{\vrule\vtop{\vbox{\hrule\kern1pt
        \hbox{\vphantom{\tt/}\thinspace{\tt#1}\thinspace}}
        \kern1pt\hrule}\vrule}\thinspace} 
\def\square{\cstok{\phantom{$.$}}} 


\def\halmos{%
  \hbox{%
    \vrule width 0.5em height 0.5em depth 0pt
  }%
}

\def\square{\halmos}

\def\fermat#1{\setbox0=\vtop{\hsize4.00pc
        \smallrm\raggedright\noindent\baselineskip9pt
        \rightskip=0.5pc plus 1.5pc #1}\leavevmode
        \vadjust{\dimen0=\dp0
        \kern-\ht0\hbox{\kern-4.00pc\box0}\kern-\dimen0}}

\def\hsizeplusepsilon{14.25truecm} 
\def\fermatright#1{\setbox0=\vtop{\hsize4.00pc
        \smallrm\raggedright\noindent\baselineskip9pt
        \rightskip=0.5pc plus 1.5pc #1}\leavevmode
        \vadjust{\dimen0=\dp0
        \kern-\ht0\hbox{\kern\hsizeplusepsilon\box0}\kern-\dimen0}}



\def\today{July 1993}
\def\quotationone{\smallrm Nils Lid Hjort}
\def\quotationtwo{\smallrm The tolerable amount of t-ness}

\def\knep{\noalign{\vskip2pt}}
\def\dellbeta{\hbox{$\dell\mu\over \dell\beta$}}
\def\dellgamma{\hbox{$\dell\mu\over \dell\gamma$}}
\def\dellxi{\hbox{$\dell\mu\over \dell\xi$}}
\def\dellsigma{\hbox{$\dell\mu\over \dell\sigma$}}

\centerline{\bigbf The exact amount of t-ness that 
	the normal model can tolerate}

\smallskip
\centerline{\bf Nils Lid Hjort } 

\smallskip
\centerline{\sl University of Oslo and University of Oxford}


\smallskip 
{{ \smallskip\narrower\noindent\baselineskip12pt 
{\csc Abstract.} 
Suppose that the normal model is used for data $Y_1,\ldots,Y_n$, 
but that the true distribution is 
a t-distribution with location and scale parameters $\xi$ and
$\sigma$ and $m$ degrees of freedom. 
The normal model corresponds to $m=\infty$. 
Using a local asymptotic framework where $m$ is allowed to 
increase with $n$ two classes of estimands are identified.
One small class, which in particular contains the functions of $\xi$ alone, 
is only affected by t-ness to the second order, and 
maximum likelihood estimation in the two- or three-parameter 
models become equivalent.
For all other estimands it is shown that if $m\ge1.458\sqrt{n}$, 
then maximum likelihood estimation using the incorrect normal model is 
still more precise than using the correct three-parameter model. 
This is furthermore shown to be true in regression models with 
t-distributed residuals. 
We also propose and analyse compromise estimators  
that in various ways interpolate between the normal and the nonnormal models. 
A separate section extends the t-ness results 
to general normal scale mixtures, in which case the tolerance
radius around the normal error distribution takes the form
of an upper bound $.3429/\sqrt{n}$ for the variance of the
scale mixture distribution.  
 
Proving our results requires somewhat nonstandard `corner asymptotics' 
since behaviour of estimators must be studied when 
the crucial parameter $\gamma=1/m$ is close to zero,
which is not an inner point of the parameter space,
and the maximum likelihood estimator of $m$ is equal to $\infty$ 
with positive probability.

\smallskip\noindent 
{\csc Key words:} \sl
choice of model,
corner asymptotics,
deliberate bias,
guarding against heavier tails, 
ignorance is strength, 
negative degrees of freedom, 
normal scale mixtures, 
parametric robustness,
robust regression, 
tolerance radius
\smallskip}} 

\bigskip
{\bf 1. Introduction and summary.} 
The most popular model for  
independent identically distributed (i.i.d.)~data 
$Y_1,\ldots,Y_n$ is to postulate normality, 
i.e.~assuming $f(y)=\phi((y-\xi)/\sigma)/\sigma$
for suitable parameters $\xi$ and $\sigma$. 
In many situations the normal density is too light-tailed
to constitute a serious description, however. 
A remedy then is to use
$$f(y,\xi,\sigma,m)=g_m\Bigl({y-\xi\over \sigma}\Bigr){1\over \sigma},$$
where $g_m(t)$ is the t-density with $m$ degrees of freedom. 
The narrower normal model corresponds to $m=\infty$,
and it is naturally felt that for large $m$
the discrepancy between normality and t-ness shouldn't matter.
One would also expect inference methods based on the 
formally incorrect normality assumption 
to work better than inference methods based on the
wider three-parameter model, for large values of $m$, 
since the estimation variability increases with the inclusion of $m$. 

This paper studies the problem of choosing between
`narrow model' and `wide model' estimators. 
Using the narrow method means introducing a certain bias 
due to incorrect modelling, whereas using the wide method 
means allowing additional parameter estimation noise.
Choosing between the estimators therefore amounts to
a statistical balancing act with perhaps deliberate bias 
against sampling variability. 

Suppose for example that the parameter to be estimated is 
the upper quartile $q=F^{-1}(.75)$. 
Concentrating on maximum likelihood 
(ML) estimators the two methods amount to
$$\hatt q_{\rm narr}=\hatt\xi_{\rm narr}+.675\,\hatt\sigma_{\rm narr}
	\quad {\rm and} \quad
  \hatt q_{\rm wide}=\hatt\xi_{\rm wide}
	+G^{-1}(.75,\hatt m_{\rm wide})\,\hatt\sigma_{\rm wide}.$$ 
Here $G(.,m)=G_m(.)$ is the t-distribution function 
with $m$ degrees of freedom, with inverse $G^{-1}(.,m)=G_m^{-1}(.)$, 
whereas $\hatt\xi_{\rm narr}$ and $\hatt\sigma_{\rm narr}$ 
are the ML estimators under the narrow two-parameter model,
and $\hatt\xi_{\rm wide}$, $\hatt\sigma_{\rm wide}$, $\hatt m_{\rm wide}$ 
are the ML estimators under the wide three-parameter model.
(The narrow model estimators are of course the ordinary 
empirical mean and empirical standard deviation statistics,
whereas ML estimates under the wide model must be computed 
by numerical maximisation techniques.)
How large must $m$ be in order for $\hatt q_{\rm narr}$ 
to be more precise than $\hatt q_{\rm wide}$? 
Suppose for a second example that the parameter to be estimated 
is ${\rm sd}$, the standard deviation for $Y_i$'s. 
We should compare
$$\hatt{\rm sd}_{\rm narr}=
	\Bigl[{1\over n}\sum_{i=1}^n(Y_i-\bar Y_n)^2\Bigr]^{1/2}
	\quad {\rm and} \quad 
\hatt{\rm sd}_{\rm wide}
	=\sqrt{{\hatt m_{\rm wide}\over \hatt m_{\rm wide}-2}}
		\,\hatt\sigma_{\rm wide}. $$
For what range of $m$ values is the narrow estimator 
more precise than the much more laborious wide estimator,
and for which $m$ values will it be advantageous to use the latter?
And a third example is that of estimating 
the probability $p(y)={\rm Pr}\{Y_i\le y\}$, 
in which case the two estimators to be compared are
$$\hatt p_{\rm narr}(y)
	=\Phi\bigl((y-\hatt\xi_{\rm narr})/\hatt\sigma_{\rm narr}\bigr)
	\quad {\rm and} \quad
  \hatt p_{\rm wide}(y)=G\bigl((y-\hatt\xi_{\rm wide})/\hatt\sigma_{\rm wide},
		\hatt m_{\rm wide}\bigr). $$

These problems have a surprisingly sharp and simple solution.
A natural large-sample framework is established in section 2,
where results are reached for the large-sample behaviour of 
ML estimators for $(\xi,\sigma)$ in the narrow model and of
ML estimators for $(\xi,\sigma,\gamma)$ in the wide model, where $\gamma=1/m$. 
These are used in section 3 to solve the problem. 
It is convenient to answer the question in terms of two classes
of estimands. For a certain small class of estimands, identified 
precisely in section 3, 
and which in particular contains all functions of the mean alone, 
t-ness turns out to be a secondary issue only, 
and `narrow' and `wide' estimation becomes large-sample equivalent. 
For the much larger class of {\it all other estimands} 
it turns out that if only $m\ge1.458\sqrt{n}$, 
then the narrow method is more precise than the wide method, 
in terms of large-sample limits of mean squared errors.
If t-ness is present with $m<1.458\sqrt{n}$ then the 
wide method is more precise. 
Thus $1.458\sqrt{n}$ is effectively the `tolerance distance' 
from the normal model w.r.t.~t-ness. 

A fuller story is told in section 4, where a general 
class of compromise estimators is considered. 
These interpolate between the narrow normal and the wider t-model.
We single out a few of these that are designed to work well both
under normal and non-normal conditions.
The regression case, where residuals can have a 
t distribution, is treated in section 5.
Results from the previous sections can be extended, 
and the tolerance distance becomes again precisely $1.458\sqrt{n}$.
Our compromise estimators in this case can be 
viewed as a basis for performing robust regression analysis,
guarding against heavier-than-normal tails.

While the primary focus of the paper is on the t-extension of the
normal to model error term distributions, 
section 6 briefly describes quite general results 
valid when the error distribution is of the normal
scale mixture type $\normal\{0,1\}/S$, where $S$ has a 
distribution suitably centred around 1. The tolerated 
amount of disturbance around the normal model takes the 
very simple form $\Var\,S\le.3429/\sqrt{n}$. 
The t-case corresponds to $S=(\chi^2_m/m)^{1/2}$, for which 
the variance is approximately equal to $1/(2m)$. 
Thus the general tolerance bound agrees with the specific t-based one.
Finally some complementing remarks are offered in section 7, 
including consequences for confidence intervals and 
discussion of situations with additional departures from 
the basic normal regression model. 

The problems about balancing modelling bias and estimation variability 
for incorrectly specified parametric models are
obviously of a general nature, 
and can be studied for other important models as well as 
on a general basis. 
Such a study is indeed reported on in Hjort (1993), 
which contains further background, 
a general theory, and explicit results 
for a generous list of commonly used statistical models.  
The present t-ness departure case 
is however non-regular and cumbersome, 
and cannot be handled as a special case of the general regular theory.
What makes this problem special is 
that the model must be studied when the 
crucial parameter $\gamma=1/m$ is close to zero, which is not
an inner point in the parameter space. 
In particular the ML estimator of $\gamma$ is zero 
(that is, the ML estimator of $m$ is infinity)
with positive probability,
the familiar ML asymptotics break down, and special methods are called for.

Classical robustness theory has been concerned in part  
with construction of inference methods that work well in certain 
{\it nonparametric} neighbourhoods of a basic model, 
for example the normal regression model. 
The present work is a contribution to 
the theory of {\it parametric} robustness, where one studies 
departures from a basic model in specific directions. 
There are perhaps two major themes in this general area:
(i) construction of new inference methods that work well both
under the idealised centre model and in the wider parametric model, and  
(ii) studies of how well or not well standard procedures 
tolerate various amounts of disturbance in these directions.
Both themes are worked on here. 

There are some robustness aspects
that are easier to study more fully in the parametric context.
The modelling bias versus estimation variance balance structure 
of model choices is more directly apparent, and one may more easily 
study general estimands. 
Our paper provides an explicit illustration of the fact that
deliberate bias can be useful, as more generally described 
by Bickel (1984) and in Hjort (1993). 
Lange, Little and Taylor (1989) studied multivariate t-distributions 
as an alternative to the normal ones, partly with the degrees of 
freedom equal to a fixed number (they report that `$m=4$ works well')
and partly treated as an unknown parameter. The latter case 
corresponds to what we call `wide estimation' here. 
Finally we point out that the idea of compromise estimators is not new.
Versions of such were considered by Bickel (1983, 1984),
and the so-called configural polysampling approach to 
robustness uses ideas similar in spirit to those of section 4 here 
to compromise between the normal and the `slash' model;
see Morgenthaler and Tukey (1991) for several papers on the subject. 
We are however able to consider completely general estimands 
with relative ease whereas the Morgenthaler and Tukey methods 
do not seem to be developed beyond the basic location and scale parameters. 

\bigskip
{\bf 2. Large sample framework for the problem.} 
The wide model has parameters $\xi$, $\sigma$, $m$. 
Let us reparameterise to $\gamma=1/m$, so that the density becomes
$$f(y,\xi,\sigma,\gamma)={c(\gamma)\over \sigma}
	\Bigl[1+\gamma\Bigl({y-\xi\over \sigma}\Bigr)
			^2\Bigr]^{-[1/2+1/(2\gamma)]}, \quad 
  c(\gamma)={\sqrt{\gamma}\over \sqrt{\pi}}
	{\Gamma(2^{-1}+(2\gamma)^{-1})\over \Gamma((2\gamma)^{-1})}.
				\eqno(2.1)$$
We are interested in this model for $\gamma$ in the vicinity of zero.
Using precise Taylor expansions and approximations to the 
$\log\Gamma(.)$ function one finds that 
$$\log f(y,\xi,\sigma,\gamma)=\log f(y,\xi,\sigma,0)
	+\gamma(\hbox{$1\over4$}z^4-\half z^2-\hbox{$1\over4$})
	-\half\gamma^2(\hbox{$1\over3$}z^6-\half z^4)
			+O(\gamma^3), \eqno(2.2)$$
in which $z=(y-\xi)/\sigma$. 
Having $\gamma=0$ corresponds to $m=\infty$ and gives back the
ordinary normal model. 

Let $\mu=\mu(f)=\mu(\xi,\sigma,\gamma)$ be some parameter estimand
of interest. We assume that $\mu$ is smooth with continuous derivatives
throughout the inner parameter space
$(\xi,\sigma,\gamma)\in(-\infty,\infty)\times(0,\infty)\times(0,\infty)$ 
and that the right derivative exists at $\gamma=0$,
$\lim_{\gamma\rightarrow0+}[\mu(\xi,\sigma,\gamma)
		-\mu(\xi,\sigma,0)]/\gamma$.
We concentrate on ML procedures, 
and wish to study the performance of the two estimators 
$$\hatt\mu_{\rm narr}=\mu(\hatt\xi_{\rm narr},\hatt\sigma_{\rm narr},0)
	\quad {\rm and} \quad
	\hatt\mu_{\rm wide}=\mu(\hatt\xi,\hatt\sigma,\hatt\gamma), \eqno(2.3)$$
where for simplicity of notation 
the subscript `wide' is dropped for the ML estimators in the
three-parameter model.

These could be compared in an 
asymptotic framework in which $Y_i$'s come from some
fixed $f(y,\xi_0,\sigma_0,\gamma)$, and $\gamma>0$. 
In this case $\sqrt{n}(\hatt\mu-\mu)$ has a limit distribution. 
The situation is different for the narrow model procedure. Here 
$\sqrt{n}(\hatt\mu_{\rm narr}-\mu)$ 
can be represented as a sum of two terms.
The first is 
$\sqrt{n}[\mu(\hatt\xi_{\rm narr},\hatt\sigma_{\rm narr},0)
	-\mu(\xi_0,\sigma_0,0)]$,
which has a limit distribution, with generally smaller 
variability than that of the wide model procedure; 
and the second is 	  
$-\sqrt{n}[\mu(\xi_0,\sigma_0,\gamma)-\mu(\xi_0,\sigma_0,0)]$, 
which tends to plus or minus infinity, reflecting a bias 
that for very large $n$ will dominate completely. 
This merely goes to show that with very large sample sizes 
one is penalised for any bias and one should use 
the wide model. 
This result is not very informative, however, and 
suggests that a large sample framework which uses
a local neighbourhood of $\gamma=0$ that shrinks 
when the sample size grows is much more adequate.  
Consider therefore model $P_n$, the $n$'th model, under which 
$$Y_1,\ldots,Y_n{\rm \ are\ i.i.d.~from\ }
	f_n(y)=f(y,\xi_0,\sigma_0,\delta/\sqrt{n}). \eqno(2.4)$$
The primary justification for working with neighbourhoods of 
exactly this size is simply that it will give informative
quantitative results.  
Here $(\xi_0,\sigma_0)$ is a fixed but arbitrary parameter point. 
The true parameter to be estimated is 
$\mu_{\rm true}=\mu(\xi_0,\sigma_0,\delta/\sqrt{n})$. 

To assess the behaviour of the estimators of $\mu$ we need 
to know what happens to narrow and wide estimators of the
respectively two and three model parameters.
Consider the score functions for the wide model,
evaluated at the null point $(\xi_0,\sigma_0,0)$.
Letting $\gamma$ carefully tend to zero in 
expressions for the three
partial log-derivatives of $f$ leads to 
$$\pmatrix{U(y) \cr
	   V(y) \cr 
	   W(y) \cr}
=\pmatrix{\dell\log f(y,\xi_0,\sigma_0,0)/\dell\xi \cr
	  \dell\log f(y,\xi_0,\sigma_0,0)/\dell\sigma \cr
	  \dell\log f(y,\xi_0,\sigma_0,0)/\dell\gamma \cr}
=\pmatrix{z/\sigma_0 \cr
	  (z^2-1)/\sigma_0 \cr
	  {1\over4}z^4-\half z^2-{1\over4} \cr}, \eqno(2.5)$$
where $z=(y-\xi_0)/\sigma_0$, cf.~(2.2).  
We shall also need the accompanying $3\times3$ size information matrix,
the covariance matrix of these three, as $Y$ has the  
$f(y,\xi_0,\sigma_0,0)$ distribution, 
i.e.~is simply $\normal\{\xi_0,\sigma_0^2\}$. One finds 
$$J_{\rm wide}={\rm VAR}_0
	\pmatrix{Z/\sigma_0 \cr 
		(Z^2-1)/\sigma_0 \cr
		{1\over4}Z^4-{1\over2}Z^2-{1\over4} \cr}
	=\pmatrix{1/\sigma_0^2 &0 &0 \cr
		  0 &2/\sigma_0^2 &2/\sigma_0 \cr
		  0 &2/\sigma_0 &7/2 \cr}. $$
Note that the upper left hand $2\times 2$ block 
$J_{\rm narr}={\rm diag}(1/\sigma_0^2,2/\sigma_0^2)$ 
is the information matrix of the narrow model, 
evaluated at $(\xi_0,\sigma_0)$.
For future reference we note that 
$$J_{\rm wide}^{-1}
	=\pmatrix{\sigma_0^2 &0 &0 \cr \knep
		0 &{7\over6}\sigma_0^2 &-{2\over3}\sigma_0 \cr \knep
		0 &-{2\over3}\sigma_0 &{2\over3} \cr},
	\quad
	J_{\rm narr}^{-1}
	=\pmatrix{\sigma_0^2 &0 \cr
		  0 & \sigma_0^2/2 \cr}. \eqno(2.6)$$

{\csc Lemma.} 
{{\sl Let $\bar U_n$ denote average of $U(Y_i)$'s, 
and similarly for $\bar V_n$ and $\bar W_n$. 
Under the sequence of models $P_n$ of (2.4),
$$\pmatrix{\sqrt{n}\,\bar U_n \cr 
	   \sqrt{n}\,\bar V_n \cr 
	   \sqrt{n}\,\bar W_n \cr}
	\rightarrow_d
  \pmatrix{0 + K \cr
	   (2/\sigma_0)\,\delta+L \cr
	   (7/2)\,\delta+M \cr}$$
as $n\rightarrow\infty$, where $(K,L,M)'\sim\normal_3\{0,J_{\rm wide}\}$.
}}	

\smallskip
{\csc Proof:} 
This follows essentially 
from the triangular version of the Lindeberg theorem.
A key observation is that 
$$f_n(y)\doteq f(y,\xi_0,\sigma_0,0)\,[1+W(y)\delta/\sqrt{n}].$$
This implies that $\bigl(U(Y_i),V(Y_i),W(Y_i)\bigr)'$ has expected value 
$\bigl(0,{2\over\sigma_0}\delta/\sqrt{n},{7\over2}\delta/\sqrt{n}\bigr)$
plus $O(n^{-1})$ terms, and that its variance matrix is 
$J_{\rm wide}+O(\delta/\sqrt{n})$. 
See also section 2 of Hjort (1993). \square 

{{\smallskip\sl
{\csc Proposition 1.}
Under model $P_n$ of (2.4) the 
behaviour of the narrow-model based estimators is given by 
$$\pmatrix{\sqrt{n}(\hatt\xi_{\rm narr}-\xi_0) \cr
	   \sqrt{n}(\hatt\sigma_{\rm narr}-\sigma_0) \cr}
	\doteq_d J_{\rm narr}^{-1}
	\pmatrix{\sqrt{n}\,\bar U_n \cr \sqrt{n}\,\bar V_n \cr}
	\rightarrow_d
	\pmatrix{0+\sigma_0^2 K \cr \sigma_0\delta+\half\sigma_0^2 L \cr},$$
in which $A_n\doteq_dB_n$ means that $A_n-B_n$ tends to zero in probability.
\smallskip}}

{\csc Proof:} 
This is essentially the familiar Taylor expansion argument,
carried out in the present local neighbourhood framework.
Note the bias term $(0,\sigma_0\delta)'$. 
The details are contained in more general arguments 
given in section 2 of Hjort (1993).
More direct methods of proof could also have been used
since $\hatt\xi_{\rm narr}$ and $\hatt\sigma_{\rm narr}$
are relatively easy to work with. \square

\smallskip
The wide method case is much more complicated 
because of the corner problem. Introduce
$$\pmatrix{A \cr B \cr C \cr}
	=J_{\rm wide}^{-1}\pmatrix{K \cr L \cr M \cr}
	\sim\normal_3\{0,J_{\rm wide}^{-1}\}. \eqno(2.7)$$
Note that $K=(1/\sigma_0^2)A$,
$L=(2/\sigma_0^2)B+(2/\sigma_0)C$. The limit in Proposition 1,
written in terms of $(A,B,C)'$, becomes 
$$\pmatrix{\sqrt{n}(\hatt\xi_{\rm narr}-\xi_0) \cr
	   \sqrt{n}(\hatt\sigma_{\rm narr}-\sigma_0) \cr}
        \rightarrow_d
	\pmatrix{0+A \cr B+\sigma_0(C+\delta) \cr}
	\sim\normal_2\{\pmatrix{0 \cr \sigma_0\delta \cr},
		\pmatrix{\sigma_0^2 & 0 \cr
			 0 & \half\sigma_0^2 \cr}\}. \eqno(2.8)$$
Since I want my reader to join me for the main story 
I defer the proof of the following proposition to the appendix.

{{\smallskip\sl
{\csc Proposition 2.} 
Under model $P_n$ of (2.4) the behaviour of the wide-model
based estimators is given by 
$$\pmatrix{\sqrt{n}(\hatt\xi-\xi_0) \cr
	   \sqrt{n}(\hatt\sigma-\sigma_0) \cr
	   \sqrt{n}(\hatt\gamma-\delta/\sqrt{n}) \cr}
	\rightarrow_d 
\cases{\pmatrix{A \cr B \cr C \cr} &if $C\ge-\delta$, \cr
       \pmatrix{A \cr B+\sigma_0(C+\delta) \cr -\delta \cr}
				       &if $C\le-\delta$. \cr} $$
}}

\bigskip
{\bf 3. Calculating the tolerance distance.}
Our programme is to use the delta method of linearisation 
in conjunction with Propositions 1 and 2 to reach limit 
distribution results for the narrow and wide estimators,
and then to compute and compare mean squared errors.

First consider the narrow method. 
Using Proposition 1 we find
$$\eqalign{
\sqrt{n}[&\mu(\hatt\xi_{\rm narr},\hatt\sigma_{\rm narr},0)
		-\mu(\xi_0,\sigma_0,\delta/\sqrt{n})] \cr
	&=\sqrt{n}[\mu(\hatt\xi_{\rm narr},\hatt\sigma_{\rm narr},0)
		-\mu(\xi_0,\sigma_0,0)]
	 -\sqrt{n}[\mu(\xi_0,\sigma_0,\delta/\sqrt{n})
		-\mu(\xi_0,\sigma_0,0)] \cr
	&\doteq_d \dellxi \sqrt{n}(\hatt\xi_{\rm narr}-\xi_0)
	 +\dellsigma\sqrt{n}(\hatt\sigma_{\rm narr}-\sigma_0)
	   	-\sqrt{n}\dellgamma\delta/\sqrt{n} \cr
	&\rightarrow_d 
	 \Lambda_{\rm narr}=\dellxi A
		+\dellsigma[B+\sigma_0(C+\delta)]-\dellgamma\delta, \cr}$$
where the partial derivatives are computed at the null model 
$(\xi_0,\sigma_0,0)$. The limit variable is normal with
$$\eqalign{
\E\Lambda_{\rm narr}&=b\delta=(\sigma_0\dellsigma-\dellgamma)\,\delta, \cr
{\rm Var}\,\Lambda_{\rm narr}
	&=\tau_0^2=[(\dellxi)^2
		+\half(\dellsigma)^2]\,\sigma_0^2. \cr}\eqno(3.1)$$
In particular the narrow method has risk 
$\E\Lambda_{\rm narr}^2=b^2\delta^2+\tau_0^2$.
See 7B for some consequences of this.  
This `narrow result' is really contained 
in general results of Hjort (1993). 

Next consider the wide method. Using Proposition 2 one finds
$$\eqalign{
\sqrt{n}[&\mu(\hatt\xi,\hatt\sigma,\hatt\gamma)
	-\mu(\xi_0,\sigma_0,\delta/\sqrt{n})] \cr
&\doteq_d \dellxi\sqrt{n}(\hatt\xi-\xi_0)
	 +\dellsigma\sqrt{n}(\hatt\sigma-\sigma_0)
	 +[(\dellgamma)+O(1/\sqrt{n})]
	  \sqrt{n}(\hatt\gamma-\delta/\sqrt{n}) \cr
&\rightarrow_d \Lambda_{\rm wide}=
  \cases{\dellxi A+\dellsigma B+\dellgamma C &if $C\ge-\delta$, \cr
	 \dellxi A+\dellsigma[B+\sigma_0(C+\delta)]-\dellgamma\delta
				&if $C\le-\delta$. \cr} \cr}$$
This is not a normal distribution. 
We calculate its mean squared error by conditioning on 
the value of $C$. Using (2.6) and (2.7) and ordinary 
techniques one finds 
$$\pmatrix{A \cr B \cr}\midd \{C=c\}
	\sim\normal_2\{\pmatrix{0 \cr -\sigma_0c \cr},
	\pmatrix{\sigma_0^2 & 0 \cr 
		 0 & \half\sigma_0^2 \cr}\}. \eqno(3.2)$$
Further calculations show that 
$$\Lambda_{\rm wide}\midd \{C=c\}\sim\cases{
	\normal\{-bc,\tau_0^2\} &if $c\ge-\delta$, \cr
	\normal\{b\delta,\tau_0^2\} &if $c\le-\delta$. \cr}$$
Accordingly $\E[\Lambda_{\rm wide}^2\midd C=c]$ 
is $b^2c^2+\tau_0^2$ if $c\ge-\delta$
and $b^2\delta^2+\tau_0^2$ if $c\le-\delta$, and 
$$\E\Lambda_{\rm wide}^2=b^2\,\E\bigl[C^2I\{C\ge-\delta\}
	+\delta^2I\{C\le-\delta\}\bigr]+\tau_0^2.$$

We are now in a position to find out when 
the narrow and risky estimator is better than the wide and safe one.
From (2.7) we can write $C=\kappa N$ 
where $\kappa^2={2\over3}$ and $N$ is normal $(0,1)$. 
From (3.1), assuming $b\not=0$, it is clear that 
the narrow method is better than the wide one if and only if 
$$\delta^2\le \E\bigl[\kappa^2N^2I\{N\ge-\delta/\kappa\}
	+\delta^2I\{N\le-\delta/\kappa\}\bigr],$$
or
$$a^2\le \E\bigl[N^2I\{N\ge-a\}+a^2I\{N\le -a\}\bigr]
	=\Phi(a)-a\phi(a)+a^2(1-\Phi(a)), \eqno(3.3)$$
using $a=\delta/\kappa$. But this is equivalent to
$0\le a\le.8399$, as borne out by numerical computations.
This means $0\le\delta\le.8399\sqrt{2/3}=.6858$,
and we have reached the following. 

\smallskip
{\csc Result.} {{\sl 
(i) The case where $b=\sigma_0\dellsigma-\dellgamma=0$ 
is rather trivial; this typically corresponds to 
a parameter estimand $\mu$ functionally independent of 
$\sigma$ and $\gamma$ at $\gamma=0$. 
In this case $\hatt\mu_{\rm wide}$ and $\hatt\mu_{\rm narr}$ 
are asymptotically equivalent, regardless of $\delta$. 
(ii) In the more interesting case
$b\not=0$, the narrow model based estimator is better than or as good
as the wider model based estimator 
(with respect to limiting mean squared error) if and only if 
$\delta\le.6858$, or $\gamma\le.6858/\sqrt{n}$,
or degrees of freedom $m\ge\sqrt{n}/.6858=1.4582\sqrt{n}$.
\smallskip}}

Lange, Little and Taylor (1989, section 3 and appendix B) 
provide information matrix calculations for the t-model
(also in the multivariate case), and case (i) of our Result
can be deduced from these. We note that the mean parameter 
$\mu=\xi$ (and more generally $\mu=x_i'\beta$ in the regression
case to be considered in section 5) has $b=0$, 
so these important estimands are not affected by moderate t-ness
at all. 

\bigskip
{\bf 4. A fuller story: compromise estimators.}
The two estimators (2.3) that have been considered so far 
have both somewhat extreme attitudes.
The first is a firm believer and the second a firm disbeliever in normality.
This section looks at some compromising methods that  
are designed to work well both under 
`close to normal' and `distinctively nonnormal' conditions.
See also section 5 in Hjort (1993). 

We have shown in Proposition 2 that $\sqrt{n}(\hatt\gamma-\delta/\sqrt{n})$
tends to $\max\{C,-\delta\}$ in distribution,
where $C\sim\normal\{0,\kappa^2\}$ and $\kappa^2={2\over3}$. 
Now shift attention to 
$T_n=\sqrt{n}\hatt\gamma/\kappa$, the natural 
statistic for testing $\gamma=0$ (normality) against 
$\gamma>0$ (t-ness). 

{{\smallskip\sl 
{\csc Theorem.}        		
Consider the general estimator
$$\mu^*=[1-w(T_n)]\,\hatt\mu_{\rm narr}
		+w(T_n)\,\hatt\mu_{\rm wide}, \eqno(4.1)$$
where $w(T_n)$ is some appropriate weight function,
assumed only to be continuous at zero 
(where the limit $T\vee0$ of $T_n$ has positive probability) 
and almost everywhere on $(0,\infty)$. Then $\sqrt{n}(\mu^*-\mu_{\rm true})$
tends in distribution to a $\Lambda$ with mean squared error 
$$\E\Lambda^2=\hbox{$2\over3$}b^2R(a)+\tau_0^2,
\quad {\sl where\ }
R(a)=\E_a\bigl[(w(T)T-a)^2I\{T>0\}+a^2I\{T\le0\}\bigr], \eqno(4.2)$$
and where $a=\delta/\kappa$ is the normalised measure of t-ness. 
\smallskip}}

{\csc Proof:}
We have 
$T_n\rightarrow_d a+\kappa^{-1}\max\{C,-\delta\}=T\vee0$,
where $T=a+C/\kappa\sim\normal\{a,1\}$, and 
$$\eqalign{
\Lambda_{\rm narr}&=b\delta+\dellxi A+\dellsigma[B+\sigma_0\kappa(T-a)], \cr
\Lambda_{\rm wide}&=
	\cases{\dellxi A+\dellsigma B+\dellgamma\kappa(T-a) &if $T>0$, \cr
	       \dellxi A+\dellsigma(B+\sigma_0\kappa T)
       		-\dellgamma a\kappa &if $T\le0$. \cr} \cr} \eqno(4.3)$$
These will next be seen to be special cases of a much more general phenomenon.
By the continuous mapping theorem
$$\sqrt{n}(\mu^*-\mu_{\rm true})
	\rightarrow \Lambda=[1-w(T\vee0)]\,\Lambda_{\rm narr}
		+w(T\vee0)\,\Lambda_{\rm wide}. $$
Some calculations lead to 
$$\Lambda\midd \{T=t\}\sim\cases
	{ \normal\{b\kappa(a-w(t)t),\tau_0^2\} &if $t>0$, \cr
	  \normal\{b\kappa a,\tau_0^2\} &if $t\le0$. \cr}$$
The claim of the theorem follows from this. \square 

\smallskip
Observe that $R(a)$ is the risk function, under squared error loss,
for the estimator $\hatt a(T\vee0)=w(T\vee0)(T\vee0)$ 
for a nonnegative parameter $a$, 
based on observing the single variable $T\vee0$, 
where $T\sim\normal\{a,1\}$. 
There is accordingly a one-to-one correspondence between 
estimators $\mu^*$ of type (4.1) for a general $\mu(\xi,\sigma,\gamma)$ 
and estimators $a^*(t)=w(t)t$ for $a$ in the structurally very simple
one-observation problem. The behaviour of any given $\mu^*$ can be
studied quite simply in terms of its associated $R(a)$ function,
and any reasonable $a$-estimator method can be transported  
to a reasonable $\mu$-estimator, via $w(t)=a^*(t)/t$.

What are interesting values of $a$? 
We have $a=\delta/\kappa$ and 
$m=1/\gamma=\sqrt{n}/\delta=\sqrt{1.5\,n}/a$,
and $T_n$ used in (4.1) detects non-normality ($m<\infty$) 
with probability $\Phi(a-1.645)$ (using level 5\%). 
This means that $T_n$ detects $a$-values beyond 4 with 
probability at least .99. 
We may think of $a$-values beyond 4, 
or $m\le .306\sqrt{n}$, as being 
t-departures from normality that should be clearly visible from data.
This tentatively suggests that estimators of the type (4.1) 
should be used with $w(t)$ close to 1 for $t\ge 4$ and
with small risk behaviour for $R(a)$ when $a\le 4$. 

A briefly annotated list of interesting choices 
for $w(T_n)$ in (4.1) follows next. 

(i) The narrow method uses $w(t)=0$, and corresponds to
using $\hatt a_{\rm narr}(t)=0$ to estimate $a$. 
Its risk is $R_{\rm narr}(a)=a^2$,
which is good for $a$ small ($m$ large)
but disastrous for $a$ large ($m$ small).

(ii) The wide method has $w(t)=1$, and corresponds to 
$\hatt a_{\rm wide}(t)=t\vee0$ to estimate $a$. Its risk is 
$$R_{\rm wide}(a)=\E_a\bigl[(T-a)^2I\{T\ge0\}+a^2I\{T\le0\}\bigr]
	=\Phi(a)-a\phi(a)+a^2(1-\Phi(a)),$$ 
cf.~(3.3). It starts at .50 at zero and climbs towards 1.
This estimator is minimax. Its risk is above .99 for $a\ge2.67$.
Again: if $a\le.8399$ then the narrow is best and if 
$a>.8399$ then the wide is best.

(iii) Try out $w(t)=w$, a constant. 
We may compute the resulting $R(a)$, and minimise w.r.t.~the choice of $w$.
The best choice, expressed in terms of the parameter point $a$, is
$$w_0(a)={a^2\Phi(a)+a\phi(a) \over
	(a^2+1)\Phi(a)+a\phi(a)}.$$
A simple idea is then to insert $T_n$ for $a$, i.e.~using
$\hatt a_{\rm ratio}=w_0(t)t$ to estimate $a$ and 
(4.1) with $w_0(T_n)$ to estimate $\mu$.
The risk $R_{\rm ratio}(a)$ starts at .249 and is better than 
$R_{\rm wide}(a)$ for $a\le1.32$, and is never much worse. 
Its maximum is 1.223, at $a=2.90$, after which it decreases towards 1. 
The narrow is better than the present one only
for $a\le.68$, and quickly becomes much worse after that.

(iv) Some natural Bayesian/empirical Bayesian ideas are as follows. 
Assume $a$ is distributed like $|\normal\{0,\tau^2\}|$,
i.e.~with prior distribution $\pi(a)={2\over\tau}\phi(a/\tau)$
on $[0,\infty)$. The Bayes solution associated with the loss function
implicit in (4.2) can be seen to be the familiar $\E\{a\midd T=t\}$ 
if $t>0$ and an arbitrary value can be used when $T\vee0=0$,
i.e.~when information on $T$ is $T\le0$.  
In the present case the Bayes solution becomes 
$$\hatt a_\tau(t)=\E\{a\midd t\}
	=\nu t+\sqrt{\nu}\phi(\sqrt{\nu}t)/\Phi(\sqrt{\nu}t),
	\quad {\rm where\ }\nu=\tau^2/(\tau^2+1).$$
Since $\E_aT=a$ and $\E a^2=\tau^2$ 
a simple empirical estimate for $\nu$ is $T^2/(T^2+1)$.
This leads to
$$\hatt a_{\rm eb}(t)={t^2\over t^2+1}t+{t\over \sqrt{t^2+1}}
	\phi\Bigl({t\over \sqrt{t^2+1}}t\Bigr)\Big/
	\Phi\Bigl({t\over \sqrt{t^2+1}}t\Bigr).$$ 
Performance: 
The risk $R_{\rm eb}(a)$ starts at .337 and is better than $R_{\rm wide}(a)$ 
for $a\le2.09$, and is never much worse. 
It is not quite as good as the narrow method when $a\le.67$,
but quickly becomes much better after that.
It reaches its maximum value of only 1.147 at $a=3.75$,
and decreases towards 1 thereafter. 

(v) The limit of the Bayes rules above, when $\tau\rightarrow\infty$,
is $\hatt a_{\rm vag}(t)=t+\phi(t)/\Phi(t)$. This is also 
the Bayes solution under a vague flat prior on the {half}{line}.
It is minimax like the $\hatt a_{\rm wide}$, but has a differently
shaped risk function, see the figure. 

(vi) Finally we could mention pre-test and related estimators.
The if-else of pre-test estimator uses 
$w(t)=0$ if $t\le d$ and $w(t)=1$ if $t>d$ in (4.1), 
and corresponds to 
$\hatt a_{\rm pre}(t)=0$ if $t\le d$ 
and $\hatt a_{\rm pre}(t)=t$ if $t>d$, for suitable cut-off value $d$. 
The theory of section 3 suggests that $d=.8399$ is a good choice, 
for example. It has risk 
$$R_{\rm pre}(a)=\Phi(a-d)+a^2[1-\Phi(a-d)]-(a-d)\phi(d-a).$$
These are inadmissible estimators in the decision-theoretic sense;
each can be uniformly improved upon (except in the border-line
cases $d=0$, which is the wide method, and $d=\infty$,
which is the narrow method). 
A related but smoother version is the limited translation variety 
$\hatt a_{\rm lim}(t)=0$ if $t\le d$ 
and $\hatt a_{\rm lim}(t)=t-d$ if $t>d$. 
This corresponds to using $w(t)=0$ if $t\le d$ and $w(t)=1-d/t$ if $t>d$.
The risk function becomes
$$R_{\rm lim}(a)=(1+d^2)\Phi(a-d)+a^2[1-\Phi(a-d)]-(a+d)\phi(a-d),$$
with maximum value $1+d^2$ occurring at infinity. 
The latter estimators are presumably reasonable approximations to 
minimax solutions subject to doing well at zero, following
the work of Bickel (1983) in a more regular situation. 

It is worth mentioning that in the general but regular case
treated in Hjort (1993), where $a=\delta/\kappa$ can vary freely
on the line, then methods (v) and (ii) above become equivalent, 
as do ideas (iii) and (iv). 

\centerline{\includegraphics[scale=0.55]{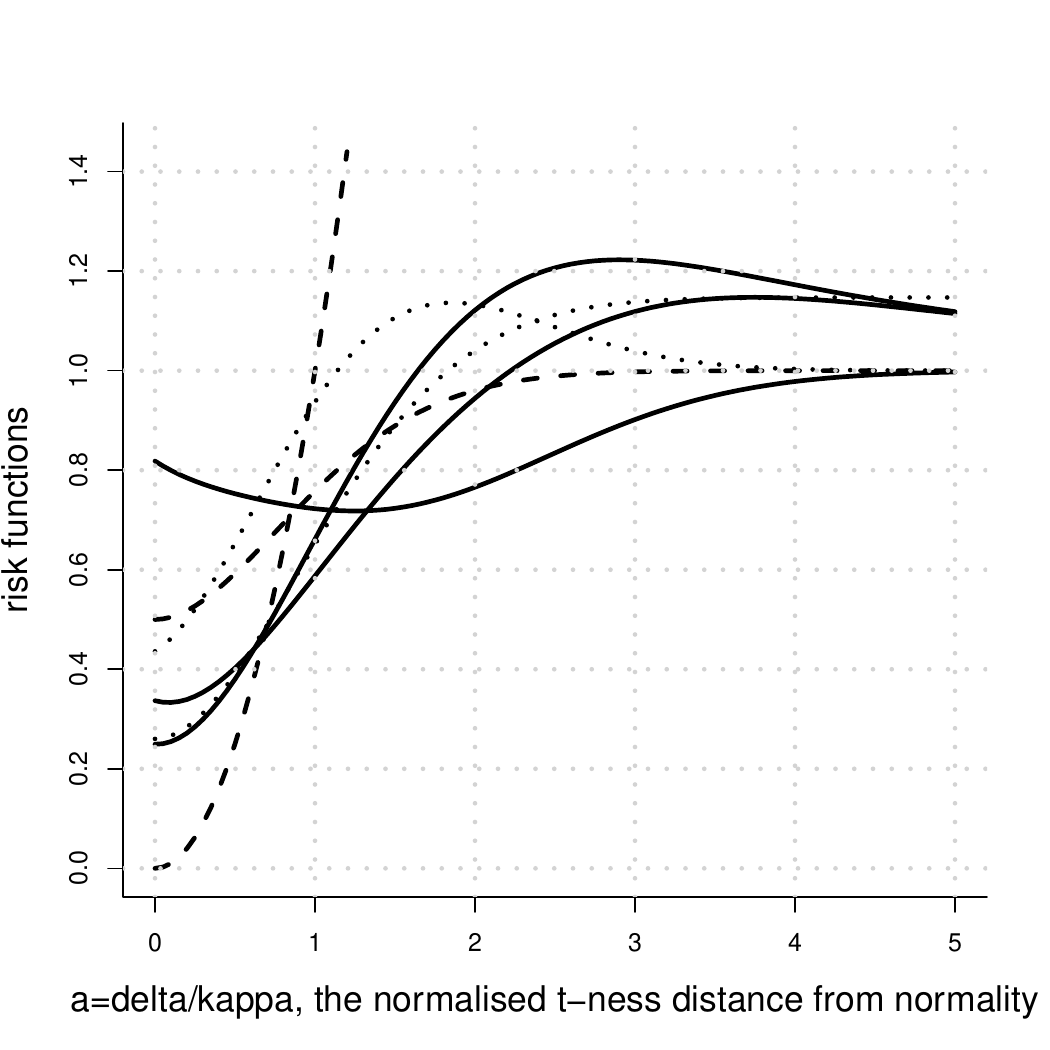}}

{{\medskip\baselineskip11pt\narrower\noindent\sl
{\csc Figure 1.} 
Risk functions $R(a)$ are shown for seven procedures, 
corresponding to seven choices of weight function $w(.)$ in (4.1). 
Risks for the wide and the narrow methods start 
at respectively .500 and .000 and are shown with dotted lines. 
The three shown with solid lines are 
the ratio method (starting at .249),
the empirical Bayes method (starting at .337), 
and the Bayes solution under uniform prior on the {half}line 
(starting at .819).
Finally shown with dashed lines are 
the pre-test method with $d$ = .8399 (starting at .436)
and the limited translation procedure with $d$ chosen to make maximum risk
$1+d^2$ equal to that of the empirical Bayes method. 
 \smallskip}}

Several risk functions $R(a)$ are plotted in the figure. 
These were computed by numerical integration where necessary. 
Overall both the ratio method (iii) and the empirical Bayes method (iv)
seem to be quite satisfactory solutions: 
they are almost as good as the wide method 
under distinctively non-normal conditions 
and are considerably better than the wide method 
under close-to-normal conditions. The limited translation procedure
of (vi) is also quite good, and is relevant if one's aim is to 
minimise the maximum risk subject to bounding the risk at the narrow model. 

In the end one should also compute $b$ and $\tau_0$ of (3.1),
or (5.2) in the regression case to be discussed, 
to see the specific consequences in terms of risk function 
(4.2) for the estimand under study; see the examples in 7A. 

\bigskip
{\bf 5. Extension to regression models.}
Suppose 
$$Y_i=x_{i,1}\beta_1+\cdots+x_{i,p}\beta_p+\sigma Z_i
	=x_i'\beta+\sigma Z_i, \eqno(5.1)$$ 
where $\beta_1,\ldots,\beta_p$ are regression parameters and 
the $Z_i$'s are i.i.d.~from a $t_m$-distribution. 
How risky are the ordinary methods, 
that all have $m=\infty$ as underlying assumption? 
How statistically noisy are the more ambitious $p+2$-parameter
methods that employ ML estimates for $\beta_1,\ldots,\beta_p,\sigma,m$? 
For example, should one use 
$$\hatt\mu_{\rm narr}=\sqrt{2/\pi}\,\hatt\sigma_{\rm narr}
	\quad {\rm or} \quad
  \hatt\mu_{\rm wide}=\sqrt{1/\pi}\,
	\bigl[\Gamma(\half\hatt m_{\rm wide}-\half)
       	/\Gamma(\half\hatt m_{\rm wide})\bigr]\,\hatt\sigma_{\rm wide}$$
to estimate $\mu=E|Y(x)-x'\beta|$, the expected distance 
from regression curve to data point? What is an effective yet safe
way of estimating regression deciles? 

Let us briefly indicate how results from earlier sections 
extend to this situation. Let $\beta_0$ and $\sigma_0$ be
arbitrary but fixed, and let $\gamma=1/m$ tend to zero 
like $\gamma=\gamma_n=\delta/\sqrt{n}$. 
The score function becomes 
$$\pmatrix{U(y_i) \cr
	   V(y_i) \cr 
	   W(y_i) \cr}
=\pmatrix{\dell\log f(y_i,\beta_0,\sigma_0,0)/\dell\beta \cr
	  \dell\log f(y_i,\beta_0,\sigma_0,0)/\dell\sigma \cr
	  \dell\log f(y_i,\beta_0,\sigma_0,0)/\dell\gamma \cr}
=\pmatrix{z_ix_i/\sigma_0 \cr
	  (z_i^2-1)/\sigma_0 \cr
	  {1\over4}z_i^4-\half z_i^2-{1\over4} \cr}, $$
in which $z_i=(y_i-x_i'\beta_0)/\sigma_0$. 
The $(p+2)\times(p+2)$ information matrix becomes
$$J_{\rm wide}=
	\lim_{n\rightarrow\infty}
	{1\over n}\sum_{i=1}^n{\rm VAR}_0
	\pmatrix{Z_ix_i/\sigma_0 \cr 
		(Z_i^2-1)/\sigma_0 \cr
		{1\over4}Z_i^4-{1\over2}Z_i^2-{1\over4}\cr}
	=\pmatrix{D/\sigma_0^2 &0 &0 \cr
		  0 &2/\sigma_0^2 &2/\sigma_0 \cr
		  0 &2/\sigma_0 &7/2 \cr}, $$
in which it is assumed that the $p\times p$ design matrix
$D$, the limit in probability of $D_n={1\over n}\sum_{i=1}^nx_ix_i'$,
exists. We note that 
$$J_{\rm wide}^{-1}
	=\pmatrix{\sigma_0^2D^{-1} &0 &0 \cr \knep
		0 &{7\over6}\sigma_0^2 &-{2\over3}\sigma_0 \cr \knep
		0 &-{2\over3}\sigma_0 &{2\over3} \cr},
	\quad
	J_{\rm narr}^{-1}
	=\pmatrix{\sigma_0^2D^{-1} &0 \cr
		  0 & \sigma_0^2/2 \cr}. $$
The parallel to section 2's Lemma is that 
$$\pmatrix{\sqrt{n}\,\bar U_n \cr 
	   \sqrt{n}\,\bar V_n \cr 
	   \sqrt{n}\,\bar W_n \cr}
	\rightarrow_d
  \pmatrix{0 + K \cr
	   (2/\sigma_0)\,\delta+L \cr
	   (7/2)\,\delta+M \cr}, $$
where $(K,L,M)'\sim\normal_{p+2}\{0,J_{\rm wide}\}$.
This is true by the triangular Lindeberg theorem 
under the usual condition 
${1\over n}\max_{i\le j\le n}(x_{i,j}-\bar x_i)^2\rightarrow0$ 
for each $i$. For the familiar normality-based 
(least-squares-type) estimators one finds 
$$\eqalign{
\pmatrix{\sqrt{n}(\hatt\beta_{\rm narr}-\beta_0) \cr
	   \sqrt{n}(\hatt\sigma_{\rm narr}-\sigma_0) \cr}
	&\doteq_d J_{\rm narr}^{-1}
	 \pmatrix{\sqrt{n}\,\bar U_n \cr \sqrt{n}\,\bar V_n \cr}
	 \rightarrow_d
	 \pmatrix{0+\sigma_0^2 K \cr\sigma_0\delta+\half\sigma_0^2 L} \cr
	&=\pmatrix{0+DA \cr B+\sigma_0(C+\delta) \cr}
	 \sim\normal_{p+1}\{\pmatrix{0 \cr \sigma_0\delta \cr},
		\pmatrix{\sigma_0^2D^{-1} & 0 \cr
			 0 & \half\sigma_0^2 \cr}\}, \cr}$$
writing $(A,B,C)'$ for $J_{\rm wide}^{-1}(K,L,M)'$, which is 
$\normal_{p+2}\{0,J_{\rm wide}^{-1}\}$. 
Next, regarding the ML estimators $\hatt\beta$, $\hatt\sigma$, $\hatt\gamma$ 
in the wider $p+2$-parameter model, 
Proposition 2 with proof can be lifted mutatis mutandis and becomes
$$\pmatrix{\sqrt{n}(\hatt\beta-\beta_0) \cr
	   \sqrt{n}(\hatt\sigma-\sigma_0) \cr
	   \sqrt{n}\hatt\gamma/\kappa \cr}
	\rightarrow_d 
\cases{\pmatrix{A \cr B \cr T \cr} &if $T\ge0$, \cr
       \pmatrix{A \cr B+\sigma_0\kappa T \cr 0 \cr}
				       &if $T\le0$. \cr} $$

The rest of the story is very similar to that of sections 3 and 4.
The limit variables $\Lambda_{\rm narr}$ and $\Lambda_{\rm wide}$
are as in (4.3) only with $(\dellbeta)'DA$
replacing $\dellxi A$, 
and $A$ is now $\normal_p\{0,\sigma_0^2D^{-1}\}$ and not 
merely $\normal\{0,\sigma_0^2\}$. 
Conditioning on $T$ one finds in the end that 
the main result (4.2) is true, with 
$$b=\sigma_0\dellsigma-\dellgamma, \quad
  	\tau_0^2=[(\dellbeta)'D^{-1}(\dellbeta)
			+\half(\dellsigma)^2]\,\sigma_0^2. \eqno(5.2)$$
Section 3's main result about $m\ge1.458\sqrt{n}$ 
for all estimands with $b\not=0$ is also true verbatim.
And for the problem of performing linear regression analysis
when the residuals could have fatter tails than the normal, 
a natural proposal is to use 
$$\mu^*=[1-w(\sqrt{1.5\,n}\,\hatt\gamma)]\,
	 \mu(\hatt\beta_{\rm narr},\hatt\sigma_{\rm narr},0)
	+w(\sqrt{1.5\,n}\,\hatt\gamma)\,
	 \mu(\hatt\beta,\hatt\sigma,\hatt\gamma), \eqno(5.3)$$
where $w(.)$ is as in (iii) or (iv) of section 4. 

\bigskip
{\bf 6. General normal scale mixtures.}
The primary focus of this article has been the t-extension
of the normal to model fatter tails in the error distribution. 
This choice is partly dictated by tradition and familiarity;
other models might more successfully fit many data sets 
with the heavier tails syndrome but everybody knows the t. 
In addition Lange, Little and Taylor (1989) and others 
report good robustness results using the t. In any case it is comforting
that results quite similar to those of previous sections can
be obtained for the large class of general normal scale mixtures, 
and that these results agree qualitatively with those for t-ness, 
as we now demonstrate.
 
Suppose $Y_i=\xi+\sigma Z_i$ where the $Z_i$'s are 
i.i.d.~and distributed as $\normal\{0,1\}/S$, 
say, where $S$ is independent of the nominator. 
Let us merely assume that $S$ has a density 
$h_\gamma(s)$ on $(0,\infty)$ and that this distribution 
becomes increasingly concentrated around 1 as $\gamma\arr0$. 
For technical convenience we stipulate that 
$\E_\gamma S=1+k_1\gamma+O(\gamma^2)$,
$\E_\gamma(S-1)^2=k_2\gamma+O(\gamma^2)$,
$\E_\gamma(S-1)^3=O(\gamma^2)$, and 
$\E_\gamma(S-1)^4=O(\gamma^2)$;
these are implied by moment convergence of $(S-1)/(k_2\gamma)^{1/2}$
to the standard normal, for example. 
Results will be seen to depend only upon the local mean 
and the local variance of $S$, i.e.~on $k_1$ and $k_2$ alone. 
The limiting case gives a unit point mass at $S=1$ which
corresponds to a perfectly normal error distribution. 
Again the programme is to consider a sequence of 
contiguous alternatives to the normal where $\gamma=\delta/\sqrt{n}$. 
The t-case hitherto considered is 
the special case where $S=(\chi^2_m/m)^{1/2}$ and $\gamma=1/m$,
and for which $k_1=-{1\over4}$ and $k_2=\half$. 

The density of $Z_i$ is 
$f_\gamma(z)=\int_0^\infty \phi(zs)sh_\gamma(s)\,\d s$.
Expand for $s$ close to 1 to find 
$$\phi(zs)s=\phi(z)
\bigl[1-(z^2-1)(s-1)+\half(z^3-3z)z(s-1)^2+O((s-1)^3)\bigr]. $$
With some further details this leads to  
$f_\gamma(z)=\phi(z)[1+\gamma R(z)+\gamma^2R_2(z)+O(\gamma^3)]$, 
where 
$$R(z)=\half k_2(z^4-6z^2+3)+k_3(z^2-1), 	
	\quad {\rm writing\ }k_3=\hbox{$3\over2$}k_2-k_1, \eqno(6.1)$$
and $R_2(z)$ is a certain polynomial of degree eight and with
mean zero under the normal. This gives 
$$\log f_\gamma(z)=\log\phi(z)+\gamma R(z)
	-\half\gamma^2S(z)+O(\gamma^3), \eqno(6.2)$$
parallelling (and generalising) (2.2) and (A.1) in the Appendix. 
Here $S(z)=R(z)^2-R_2(z)$,
and in the final analysis (see the t-case proof in the Appendix) 
it only matters that $S(Z)$ has mean
equal to the variance of $R(Z)$ in the null case. 

One can now go through the chains of reasoning from previous sections
and obtain analogue results for the present general model. 
There is a new $J_{\rm wide}$ matrix, and the analogue of 
Proposition 2 features $B+k_3\sigma_0(C+\delta)$, 
where $(A,B,C)'\sim\normal_3\{0,J_{\rm wide}^{-1}\}$  
(and $k_3=1$ for the t-case). We omit details, but report that 
the narrow and the wide limiting risks are 
$$\eqalign{
\E\Lambda_{\rm narr}^2&=(b^*\delta)^2+\tau_0^2, \cr
\E\Lambda_{\rm wide}^2&=(b^*)^2\E[C^2I\{C\ge-\delta\}
	+\delta^2I\{C\le-\delta\}]+\tau_0^2, \cr}$$
featuring a new $b^*=k_3\sigma_0\dellsigma-\dellgamma$,
while $\tau_0^2$ still is as in (3.1).  
The variance $\kappa^2$ of $C$ is now $1/(6k_2^2)$. 
As in sections 3 and 4 it is convenient to use $a=\delta/\kappa$ 
as a measure of normalised distance from normality. 
The arguments of section 3 lead to the following. 

{{\smallskip\sl
{\csc General Result.} 
(i) If the $\mu=\mu(\xi,\sigma,\gamma)$ estimand is such that
$b^*=0$, then $\hatt\mu_{\rm wide}$ and $\hatt\mu_{\rm narr}$
are asymptotically equivalent, regardless of $\delta$. 
(ii) If the $\mu$ estimand is such that $b^*\not=0$, 
then narrow model based estimation is better than or as good as 
wide model based estimation (with respect to limiting mean 
squared error) if and only if $a\le.8399$, 
or $\gamma\le.8399\kappa/\sqrt{n}$, or 
$k_2\gamma=\Var\,S\le.3429/\sqrt{n}$.
\smallskip}}

Exactly the same tolerance radius is found in the regression
context $Y_i=x_i'\beta+\sigma Z_i$ when $Z_i$ is distributed 
as a standard normal divided by $S$, after parallelling the work
of section 5. The answer is perhaps surprisingly simple,  
involving only the variance of the scale mixture, but comes about
since our large-sample framework only is concerned with
local alternatives to the normal. The general result also 
indicates that conclusions of previous sections based on the t-ness model, 
including the compromise recipes of section 5, 
remain qualitatively correct in a broader robustness context. 


There are also parametric families with both heavier and lighter
than normal tails. 
In the technical report version of this article calculations are
given for a quasi-extension of the t allowing negative degrees of
freedom. 
A better example is the exponential power family 
discussed in Box and Tiao (1983, section 3.2). Calculating
the tolerance distance and devising compromise estimators 
is much easier than for the t-case, applying general methods of Hjort (1993), 
due to the fact that the normal in these cases is 
an inner member of the family. 
Finally we point out that there are other
natural extensions of the basic normal model 
that also involve problems with corners of parameter spaces, 
and where methods of this paper should be useful. 
One example is the normal contamination model. 

\bigskip
{\bf 7. Additional remarks.}

{\sl 7A. Some estimands.} 
To illustrate both the general formulae and the relative importance
of bias and estimation noise, let us go through a short list 
of important estimands. 

(i) Let $\mu=x'\beta$, the regression curve at a specific point.
Then $b=0$ and all compromise estimators become asymptotically equivalent,
with $\tau_0^2=\half x'D^{-1}x\,\sigma_0^2$ as limiting normalised risk.
Thus familiar least-squares estimates are sufficiently precise 
even in the presence of t-ness, and the same is true in 
other cases where the estimand only depends upon $\beta_1,\ldots,\beta_p$.

(ii) Let $\mu=E|Y(x)-x'\beta|$, our starting example of section 5.
Then $\mu=\sigma\,E|Z|$, where $Z$ is $t_m$-distributed, and 
some determined analysis gives 
$b=\half\sigma_0\phi(0)$, $\tau_0^2=2\phi(0)^2\sigma_0^2$.
This gives 
$${\rm risk}={\sigma_0^2\over \pi}\Bigl[{1\over 12}R(a)+1\Bigr]$$
for the limit distribution version of $n$ times mean squared error
for $\mu^*$, see (4.2). 

(iii) Let $\mu$ be the $p$-th quantile of the distribution for 
$Y(x)$ at $x$. It is for example often useful and illuminating 
to draw the nine regression deciles (corresponding to $p=j/10$) 
in the same diagram, as functions of $x$. 
Then $\mu=x'\beta+\sigma G^{-1}(p,m)$ in the notation of section 1.
One can work out a suitable expression for $\dell\mu/\dell\gamma$,
and then find $b$ and $\tau_0$ of (5.2). The end result is
$${\rm risk}=\bigl\{\hbox{$2\over3$}[z_p+A(z_p)/\phi(z_p)]^2R(a)
	+x'D^{-1}x+\half z_p^2\bigr\}\,\sigma_0^2,$$
in which $z_p=\Phi^{-1}(p)$ and 
$A(t)=\int_{-\infty}^t \phi(z)W(z)\,dz$, and $W(z)$ is as in (2.5).
Formula (5.3), with a weight function deemed appropriate in view
of the discussion of section 4, provides a regression quantile estimator
that is robust against fatter than normal tails of the error distribution.
More generally robust methods for regression quantiles are in 
Koenker and Bassett (1978), Koenker and Portnoy (1987), 
and in Hjort and Pollard (1993, section 3D). 

(iv) The case of a probability $\mu={\rm Pr}\{Y(x)\le y\}
=G\bigl((y-x'\beta)/\sigma,m\bigr)$ is similar to but simpler than 
case (iii). The same expression for risk emerges, with
$z(y)=(y-x'\beta_0)/\sigma_0$ replacing $z_p$.


\subsection
{\sl 7B. False confidence.}
We proved in section 3 that 
$\sqrt{n}(\hatt\mu_{\rm narr}-\mu_{\rm true})$ tends to 
$\normal\{b\delta,\tau_0^2\}$ under the (2.4) sequence of models.
Traditional normality-based inference 
uses in essence that the limit is $\normal\{0,\tau_0^2\}$.
Accordingly $b^2\delta^2$ is the 
invisible extra burden associated with using 
the normality-based estimator 
when in fact the wider model (2.4) is true. 
Consequences of this include that 
traditional normality-based confidence intervals and testing
procedures behave incorrectly; the intervals have adequate length 
but are incorrectly placed, 
and the tests have too high significance levels. If 
${\rm CI}_{\rm narr}=\hatt\mu\pm1.645\hatt\tau_0/\sqrt{n}$,
for example, then the coverage probability converges to 
${\rm Pr}\{|\normal\{b\delta/\tau_0,1\}|\le1.645\}$,
which is strictly less than 90\% unless $b=0$ or $\delta=0$. 
See also section 4H of Hjort (1993). 

\subsection
{\sl 7C. How far away is $1.458\sqrt{n}$?}
One can test normality $(\gamma=0)$ against t-ness $(\gamma>0)$
using $T_n=\sqrt{1.5\,n}\,\hatt\gamma$, see section 4.
The limit distribution under normality is $\max\{N,0\}$ where 
$N$ is standard normal. The $T_n>1.645$ test has 
(asymptotic) level 5\% and power $\Phi(a-1.645)$. 
One way of quantifying the 
distance from normality to the first intolerable t-distribution 
is in terms of $\Phi(0.8399-1.645)=.210$, the probability of 
detecting this amount of t-ness. The corresponding detection probability 
figure is .329 for the case of a 10\% level test. 
These rather low power figures, even for the strong test that
looks specifically for t-deviation, 
suggest that $1.458\sqrt{n}$ is not very far away from the normal,
and that this moderate amount might not be easily visible.
Statisticians often look at quantile-quantile plots to check
approximate normality, and simulations demonstrate that these plots often 
look fairly linear with this amount of t-ness. 
As a little experiment, consider the natural quantile test statistic 
$$D_n=\max_i\sqrt{n+2}
\big|Y_{(i)}-\hatt\xi_{\rm narr}-\hatt\sigma_{\rm narr}\Phi^{-1}(i/(n+1))\big|$$
for normality against non-normal tails, where the maximum is over 
$.025\le i/(n+1)\le.975$, say 
(it is scaled in such a way that it has a limit distribution in
terms of a zero-mean Gau\ss ian process);
this is the formal version of looking for non-straightness of the 
quantile plot. The power of the 5\% version of this 
test for $n=100$ at the suggested tolerance limit with 
14.58 degrees of freedom is about 13\%. The corresponding power figure
is 23\% for a 10\% level test. 

The bottom line is that standard normal-based procedures for 
estimands other than the simple mean parameter (more precisely,
those where $b$ of (3.1) or (5.2) is not zero) are in danger. 
The risk calculations of section 4 and the confidence undershooting  
pointed out in 7B are relevant for quite moderate doses of fat tails. 
This agrees also with the general scale mixture calculations of
section 6. Remedies include full estimation in the wide model, which is 
what Lange, Little and Taylor (1989) do, or compromise as with (5.3). 

There are other measures of distance from normality to t-ness 
that could fruitfully be used here; 
see Hjort (1993, sections 4B and 4C) 
for other proposals and interpretations in a general parametric
robustness context. 
The $L_1$-distance $\int |f_\gamma-f_0|\,\d y$ is approximately 
$.434/\sqrt{n}$ at the tolerance threshold, for example. 
Kass (1989, p.~211) mentions as a piece
of statistical trivium that the t-density which in the sense
of a certain curvature measure lies half-way between the 
Cauchy and the normal has $m=3$ degrees of freedom, 
and some calculations give that on the $L_1$-distance scale 
the corresponding answer has $m=2.3$ degrees of freedom. 
The point is again that $m\ge1.458\sqrt{n}$ means being 
reasonably close to normality. 

\subsection
{\sl 7D. Additional deviances from the normal.}
In many situations one would have to worry about additional 
ways in which the normal regression model could be wrong,
for example in the direction of variance heterogeneity or 
nonlinearity of the mean. This calls for an extension of our
framework to moderate misspecification in 
several directions, where the first is t-ness. 
Our results can be suitably generalised and become more nuanced. 
There will be a sacrosanct region around the normal inside of which 
the precision of {\it each normality-based estimator} 
is at least as good as that of the corresponding wide model estimator. 
{\it Each given estimand} will however have its own, 
larger tolerance region where narrow is more precise than wide estimation. 
See Hjort (1993, section 5I). 

\subsection
{\sl 7E. Other loss functions.}
Our risk functions and tolerance limits have been computed 
starting from the squared error criterion, where squared bias 
and variance are exchangeable currencies. Similar results can
be reached for other loss functions, but become more complicated.
If absolute loss $\sqrt{n}|\mu^*-\mu_{\rm true}|$ is used,
then there is again a one-one correspondence between estimators 
(4.1) of $\mu$ and estimators
$\hatt a(t)=w(t)t$ of the non-negative $a$ 
based on $T\vee 0$ where $T\sim\normal\{a,1\}$, 
but the absolute loss for $\mu$
transforms into the different loss function 
$\int_0^\infty L(\rho(\hatt a(t)-a))\phi(t-a)\,\d t$ for $a$,
where $L(z)=\E|z+\normal\{0,1\}|=z+2\phi(z)-2[1-\Phi(z)]$,
and $\rho=|b|\kappa/\tau_0$. The tolerance limit now depends 
on the parameter estimand via the value of $\rho$. 
See also Hjort (1993, section 5H). 

\subsection
{\sl 7F. A quasi-extension of the t-distribution
with negative degrees of freedom.}
It was necessary to use non-standard corner asymptotics to reach 
results in sections 3--5. The problems would have been much simpler 
to solve if the parameter space for $\gamma=1/m$ had
included zero as an inner point, 
i.e.~if the model had permitted negative values of $\gamma$. 
This is not only a technical but also a statistical point, 
since data sets could easily display lighter-than-normal tails
(negative kurtosis), and in a way it is an artificial facet
of the smooth transition from t-ness to normality 
(letting $m\rightarrow\infty$) that it has stop right there.

It is therefore tempting to by-pass the whole t-model 
and create a new alternative model $f(y,\xi,\sigma,\gamma)$ 
that permits negative values of $\gamma$. 
Inspired by (2.2) one could try 
$$f(y,\xi,\sigma,\gamma)=\phi\Bigl({y-\xi\over \sigma}\Bigr){1\over \sigma}
	\Bigl[1+\gamma A\Bigl({y-\xi\over \sigma}\Bigr)\Bigr]$$
for suitable $A(z)$-function. Natural desiderata are
(i) $A(z)$ is symmetric about zero;
(ii) the model is defined 
for $\gamma$'s in an interval around zero;
(iii) the density decreases with $y$ for $y\ge\xi$;
(iv) the kurtosis is positive for $\gamma>0$ and 
negative for $\gamma<0$. 

\centerline{\includegraphics[scale=0.55]{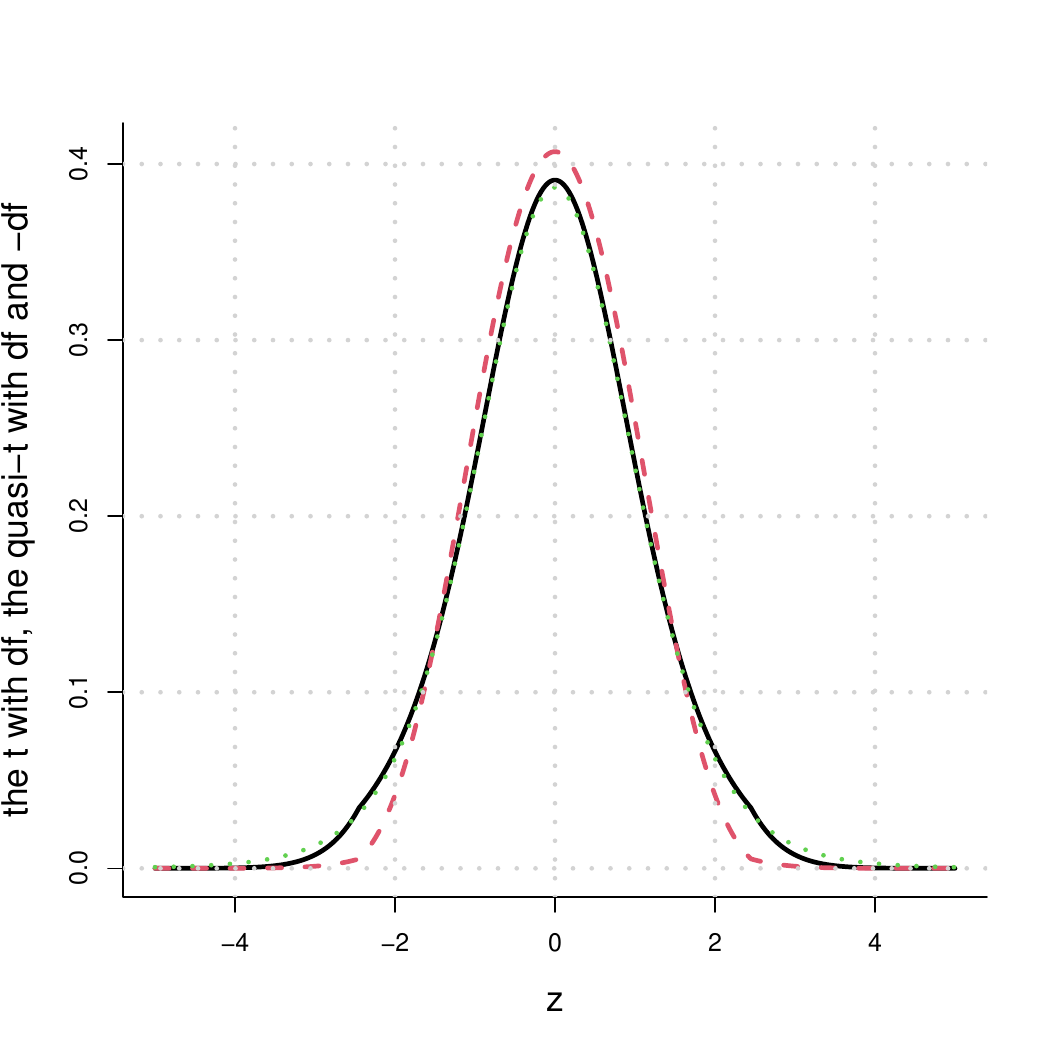}}

{{\medskip\baselineskip11.5pt\narrower\noindent\sl
{\csc Figure 2.} 
Three probability densities are shown:
(i) the quasi-t with $\df=8$, positive kurtosis, black full line;
(ii) the quasi-t with $\df=-8$, negative kurtosis, red slanted line;
and (iii) the usual t with $\df=8$, green dotted, bare
distinguishable from the quasi-t with the same $\df$.\smallskip}}

This can be achieved in various ways. Note that $A(z)$ has to be bounded 
because of (ii). Having a quasi-extension of the t-distribution 
in mind we try
$$A(z)=\cases{{1\over4}z^4-\half z^2-a(c) &if $|z|\le c$, \cr
	\noalign{\vskip2pt}
	      {1\over4}c^4-\half c^2-a(c) &if $|z|\ge c$, \cr}$$
where
$$ a(c)={1\over4}-(\half c^3+\half c)\phi(c)
+(\half c^4-c^2-\half)[1-\Phi(c)]  $$
ensures the necessary $\int\phi(z)A(z)\,dz\allowbreak=0$. 
(With some extra efforts the family could be smoothed 
at the cut-off points $\pm c$.) 
Elaborate analysis shows that (ii)--(iv) hold 
for $\gamma$-values in a suitable $(l(c),r(c))$ 
interval around zero, at least when $c\ge\sqrt{6}=2.4495$.
I have found formulae for the various necessary quantities 
($J_{\rm wide}$ etc.)~in terms of $c$. 
If $c$ is chosen large then $l(c)$ closes in on zero,
so we might as well choose $c=\sqrt{6}$,
for which the permissible symmetric interval is $(-.165,.165)$.
This defines a quasi-t-distribution with degrees of freedom
$m$ permitted to go from about 7 to infinity and over the top down to 
about $-7$. Figure 2 shows the quasi-t with $8$ and $-8$ degrees of 
freedom. The quasi-t and the t are almost identical when $m\ge10$
(you would need about 1,186 data points to be 95\% sure that 
my sequence is from the quasi-t$_{10}$ and not the usual t$_{10}$).
%
The kurtosis curve has derivative $1.244$ at $\gamma=0$ 
for this quasi-t family of probability densities,
and further analysis (but with no corner asymptotics required)
shows that the normal model can tolerate deviation up to 
$|\gamma|\le1.895/\sqrt{n}$. 

There are also other parametric families with both heavier and lighter
than normal tails. One example is the exponential power family 
discussed in Box and Tiao (1983, section 3.2). Again calculating
the tolerance distance and devising compromise estimators 
is much easier than for the t-case,
using the methods of Hjort (1993), 
since the normal is an inner member of the family.  

\bigskip
{\bf Appendix: Proof of Proposition 2.} 
By (2.2) the log-likelihood function can be written 
$$\eqalign{
L_n(\xi,\sigma,\gamma)
  =&\sum_{i=1}^n\log f(y_i,\xi,\sigma,\gamma)
	=-n\log(2\pi)^{1/2}-n\log\sigma
	 -\half\sum_{i=1}^n(y_i-\xi)^2/\sigma^2 \cr
	&+\gamma\sum_{i=1}^nR\bigl((y_i-\xi)/\sigma\bigr)
	 -\half\gamma^2\sum_{i=1}^nS\bigl((y_i-\xi)/\sigma\bigr)
	 +O_p(n\gamma^3), \cr} \eqno({\rm A.1})$$
in which 
$$R(z)=\hbox{$1\over4$}z^4-\half z^2-\hbox{$1\over4$}
	\quad {\rm and} \quad 
  S(z)=\hbox{$1\over3$}z^6-\half z^4.$$
The limit in probability of ${1\over n}L_n(\xi,\sigma,\delta/\sqrt{n})$,
under sequence (2.4), is seen to be 
$-\log(2\pi)^{1/2}\allowbreak-\log\sigma
 -\half[(\xi-\xi_0)^2+\sigma_0^2]/\sigma^2$,
uniformly over compact sets.
It follows that the sequence of ML estimators $\hatt\xi$, $\hatt\sigma$ 
must converge in probability to the parameter values that 
maximise this limit, i.e.~to the underlying $\xi_0$, $\sigma_0$. 
By working with the $\gamma$-related part of (A.1) one shows similarly
that $\hatt\gamma=\hatt\delta/\sqrt{n}$ must converge to zero
in probability.

Let in what follows $I_n(\xi,\sigma,\gamma)$ be the $3\times3$ matrix 
with elements 
$\sum_{i=1}^n(\dell^2/\dell\xi^2)\log f(y_i,\allowbreak \xi,\sigma,\gamma)$
{\it\&}cetera. If $\tilda\xi_n$, 
$\tilda\sigma_n$,  
$\tilda\gamma_n$ tend to respectively $\xi_0$, $\sigma_0$, $0$
in probability, still under the (2.4) sequence of models,  
then $-{1\over n}I_n(\tilda\xi_n,\tilda\sigma_n,\tilda\gamma_n)
\rightarrow_pJ_{\rm wide}$. This holds since direct inspection shows 
$$-{1\over n}I_n(\tilda\xi_n,\tilda\sigma_n,\tilda\gamma_n)
	=-{1\over n}I_n(\xi_0,\sigma_0,0)
	 +O_p(|\tilda\xi_n-\xi_0|+|\tilda\sigma_n-\sigma_0|+\tilda\gamma_n),$$
and the first term here can be shown to converge to $J_{\rm wide}$, 
under (2.4), using ordinary methods.

There are two possibilities regarding the maximisers of (A.1). 
Either data $(y_1,\ldots,y_n)$ are such that maximum occurs for 
some $\hatt\gamma>0$, or it occurs for $\hatt\gamma=0$. 
In the first case the ML values are solutions to 
$\dell L_n/\dell\xi=0$, $\dell L_n/\dell\sigma=0$, $\dell L_n/\dell\gamma=0$,
and the familiar Taylor argument yields 
$$\pmatrix{\sqrt{n}(\hatt\xi-\xi_0) \cr
	   \sqrt{n}(\hatt\sigma-\sigma_0) \cr
	   \sqrt{n}(\hatt\gamma-0) \cr}
  =\bigl[-I_n(\tilda\xi,\tilda\sigma,\tilda\gamma)/n\bigr]^{-1}
	\pmatrix{\sqrt{n}\,\bar U_n \cr
		 \sqrt{n}\,\bar V_n \cr
		 \sqrt{n}\,\bar W_n \cr}$$
for suitable $(\tilda\xi,\tilda\sigma,\tilda\gamma)$ 
somewhere between $(\xi_0,\sigma_0,0)$ and 
$(\hatt\xi,\hatt\sigma,\hatt\gamma)$, see the definition in (2.5). 
In the second case $L_n(\xi,\sigma,\gamma)$ 
decreases in $\gamma\ge0$, and the ML estimators 
are $(\hatt\xi_{\rm narr},\hatt\sigma_{\rm narr},0)$. 
Let $\Omega_n$ be the set of $(y_1,\ldots,y_n)$ for which 
the first case happens. Then 
$\sqrt{n}(\hatt\xi-\xi_0,\hatt\sigma-\sigma_0,\hatt\gamma-0)'$ becomes 
$$J_{\rm wide}^{-1}
	\pmatrix{\sqrt{n}\,\bar U_n \cr
		 \sqrt{n}\,\bar V_n \cr
		 \sqrt{n}\,\bar W_n \cr}+O_p(n^{-1/2})
	\quad {\rm \ or} \quad 
 \pmatrix{J_{\rm narr}^{-1}
	\pmatrix{\sqrt{n}\,\bar U_n \cr 
		 \sqrt{n}\,\bar V_n \cr} + O_p(n^{-1/2}) \cr 
						0 \cr} $$
as respectively $\Omega_n$ is in command or not. 
It turns out that $\Omega_n$ happens or not according to 
whether 
$$\Delta_n=-\hbox{$2\over3$}\sigma_0\sqrt{n}\,\bar V_n
	+\hbox{$2\over3$}\sqrt{n}\,\bar W_n>0 {\rm \ or\ }\le0, $$
to a first order approximation. 
The $\Delta_n$ here is the third component of 
$J_{\rm wide}^{-1}\sqrt{n}(\bar U_n,\bar V_n,\allowbreak\bar W_n)'$,
and the precise statement is that $I(\Omega_n)-I\{\Delta_n>0\}$
goes to zero in probability under the (2.4) regime of models. 
Using this result, the Lemma, and (2.7)--(2.8) in tandem 
yields the statement of Proposition 2, 
by the continuous mapping theorem on 
$\sqrt{n}(\bar U_n,\bar V_n,\bar W_n)'$.

To prove that $\Omega_n$ and $\{\Delta_n>0\}$ are asymptotically
equivalent events, consider once more the second half of (A.1), 
which can be expressed as 
$$\delta{1\over \sqrt{n}}\sum_{i=1}^nR\bigl((y_i-\xi)/\sigma\bigr)
	-\half\delta^2{1\over n}\sum_{i=1}^nS\bigl((y_i-\xi)/\delta\bigr) $$
plus a remainder of size $O_p(\delta^3/\sqrt{n})$. 
This is a parabola in $\delta\ge0$, 
with maximum occurring to the right of zero or at zero
depending upon the sign of the $R$-average 
(the $S$-average will be positive with probability tending to one
in the parameter region of interest). Accordingly,
apart from an $O_p(n^{-1/2})$ term,  
$$\sqrt{n}\hatt\gamma=\hatt\delta
={1\over \sqrt{n}}\sum_{i=1}^nR\bigl((y_i-\hatt\xi)/\hatt\sigma\bigr)
\Big/{1\over n}\sum_{i=1}^nS\bigl((y_i-\hatt\xi)/\hatt\sigma\bigr)$$
provided nominator is positive,
and $\hatt\delta=0$ otherwise. But 
$$\eqalign{
{1\over n}\sum_{i=1}^nR\Bigl({y_i-\hatt\xi\over \hatt\sigma}\Bigr)
&={1\over n}\sum_{i=1}^n
	R\Bigl(z_i-{\hatt\xi-\xi_0\over \sigma_0}
  	-{y_i-\xi_0\over \sigma_0^2}\,
    	(\hatt\sigma-\sigma_0)\Bigr)+O_p(n^{-1}) \cr
&={1\over n}\sum_{i=1}^nR(z_i)
   	-{1\over n}\sum_{i=1}^nR'(z_i){\hatt\xi-\xi_0\over \sigma_0}
   	-{1\over n}\sum_{i=1}^nR'(z_i)z_i\,
	{\hatt\sigma-\sigma_0\over \sigma_0}+O_p(n^{-1}), \cr}$$	
where $z_i=(y_i-\xi_0)/\sigma_0$, 
and similarly for the $S$-function term. 
Judicious calculations based on this show that 
${1\over n}\sum_{i=1}^nS\bigl((y_i-\hatt\xi)/\hatt\sigma\bigr)$ 
goes to $7/2$ and that 
$$\eqalign{
{1\over \sqrt{n}}\sum_{i=1}^nR\bigl((y_i-\hatt\xi)/\hatt\sigma\bigr)
	&=\sqrt{n}\,\bar W_n-(2/\sigma_0)\sqrt{n}(\hatt\sigma-\sigma_0)
		+O_p(n^{-1/2}) \cr
	&=\hbox{$7\over3$}\sqrt{n}\,\bar W_n
	   	-\hbox{$7\over3$}\sigma_0\sqrt{n}\,\bar V_n
					+O_p(n^{-1/2}) \cr}$$
on the set $\Omega_n$. This finally means that 
$\sqrt{n}\hatt\gamma=\Delta_n+O_p(n^{-1/2})$ 
in the $\Delta_n+O_p(n^{-1/2})>0$ case
and is $0$ in the $\Delta_n+O_p(n^{-1/2})\le0$ case. 
This proves what was needed. \square

\bigskip

\centerline{\bf References}

\medskip
\parindent0pt
\parskip3pt
\baselineskip11pt

\ref{%
Bickel, P.J. (1983). 
Minimax estimation of the mean of a normal distribution 
subject to doing well at a point.
{\sl Recent Advances in Statistics,
Festschrift for Herman Chernoff},
editors Rizvi and Siegmund, 511--528.
Academic Press, New York.}

\ref{%
Bickel, P.J. (1984).
Parametric robustness: small biases can be worthwhile.
{\sl Annals of Statistics}~{\bf 12}, 864--879.}

\ref{%
Box, G.E.P.~and Tiao, G.C. (1973).
{\sl Bayesian Inference in Statistical Analysis.}
Addison-Wesley, Menlo Park.}

\ref{%
Hjort, N.L. (1993).
Estimation in moderately misspecified models.
Submitted for publication.} 

\ref{%
Hjort, N.L.~and Pollard, D. (1993).
Asymptotics for minimisers of convex processes.
Submitted for publication.}

\ref{%
Kass, R.E. (1989).
The geometry of asymptotic inference [with discussion contributions].
{\sl Statistical Science} {\bf 3}, 188--234.}

\ref{%
Koenker, R.W.~and Bassett, G.W. (1978).
Regression quantiles.
{\sl Econometrica} {\bf 46}, 33--50.}
 
\ref{%
Koenker, R.W.~and Portnoy, S. (1987).
L-estimation for linear models.
{\sl Journal of the American Statistical Association}~{\bf 82}, 851--857.}

\ref{%
Lange, K.L., Little, R.J.A., and Taylor, J.M.G. (1989).
Robust statistical modeling using the t distribution.
{\sl Journal of the American Statistical Association}~{\bf 84}, 881--896.}

\ref{%
Morgenthaler, S.~and Tukey, J.W.~(eds) (1991).
{\sl Configural Polysampling: A Route to Practical Robustness.}
Wiley, New York.}


\bye